\documentclass[prl,aps,twocolumn,superscriptaddress]{revtex4-1}
\usepackage{amsmath}
\usepackage{color}
\usepackage{bbm}
\usepackage{amssymb}
\usepackage{epsfig}
\usepackage{multirow}
\usepackage{amsbsy}
\usepackage{array}
\usepackage{diagbox}
\usepackage{bm}
\usepackage{extarrows}
\usepackage{graphicx}
\usepackage{appendix}
\usepackage{txfonts}
\usepackage[colorlinks=true,linkcolor=blue,citecolor=blue,urlcolor=blue,bookmarks=false]{hyperref}

\begin{document}
\title{Flat and Correlated Plasmon Bands in Graphene/$\alpha$-RuCl$_3$ Heterostructures}
\author{Hui-Ke Jin}
\affiliation {Department of Physics TQM, Technische Universit\"{a}t M\"{u}nchen, $\&$ James-Franck-Straße 1, D-85748 Garching, Germany}
\author{J. Knolle}
\affiliation {Department of Physics TQM, Technische Universit\"{a}t M\"{u}nchen, $\&$ James-Franck-Straße 1, D-85748 Garching, Germany}
\affiliation{Munich Center for Quantum Science and Technology (MCQST), 80799 Munich, Germany}
\affiliation{Blackett Laboratory, Imperial College London, London SW7 2AZ, United Kingdom}

\date{\today}
\begin{abstract}
We develop a microscopic theory for plasmon excitations of graphene/$\alpha$-RuCl$_3$ heterostructures. Within a Kondo-Kitaev model with various interactions,
a heavy Fermi liquid hosting flat bands emerges in which the itinerant electrons of graphene effectively hybridize with the fractionalized fermions of the Kitaev quantum spin liquid.
We find novel correlated plasmon bands induced by the interplay of flat bands and interactions 
and argue that our theory is consistent with  the available experimental data on graphene/$\alpha$-RuCl$_3$ heterostructures. We predict novel plasmon branches beyond the long-wavelength limit and discuss the implications for probing correlation phenomena in other flat band systems. 
\end{abstract}

\maketitle

{\em Introduction.---}\label{sec:intro}
Plasmons are collective charge oscillations whose properties are normally dominated by the long-range Coulomb interactions in low density systems~\cite{Pines1952,Pines1956}. However, strong correlations can drastically alter their behavior which allows to probe new quantum many-body physics with optical experiments.  
For example, Kondo interactions can give rise to new low energy plasmon modes in heavy fermion materials~\cite{Lee1987,Millis1990,Keller1990} or can distort the surface collective modes in topological Kondo insulators~\cite{Galitski2014}. A local Hubbard interaction also leads to a strong renormalization of the plasmon dispersion and a shift of spectral weight~\cite{Loon2014,Greco2016,yin2019quantum}. Apart from strong interactions, it is of course the form of the electronic bandstructure which determines the properties of plasmons. For example, monolayer graphene serves as an outstanding platform for the study of Dirac plasmons~\cite{Wunsch2006,Hwang2007,Bonaccorso2010,Aavouris2010,Ju2011,Koppens2011,Grigorenko2012,Koppens2012,Stauber2014} with a low-energy and long-wavelength dispersion, $\omega\propto{}\sqrt{q}$, and the van Hove singularity of the dispersion leads to so-called $\pi$ plasmons which have been observed in monolayer graphene with electron energy loss spectroscopy~\cite{Eberlein2008,Kinyanjui2012,Stauber2010_1}.    

The advent of two-dimensional (2D) heterostructures has paved the
way for investigating new correlation and bandstructure effects on plasmon modes. For instance, twisted bilayer graphene (TBG), the moir\'e material which hosts strongly correlated flat bands~\cite{Cao20181,Cao20182}, shows novel collective plasmon excitations~\cite{Stauber2016,Lewandowski2019,Stauber2020,Novelli2020, Hesp2019,Hu2017,Fahimniya2020}. Apart from the conventional 2D Dirac plasmons which are damped as momentum $q$ increases and merge with the particle-hole (p-e) continuum~\cite{Wunsch2006,Hwang2007}, it is reported that the plasmons in TBG and other narrow-band materials exhibit flat and weakly damped dispersions piercing through the p-e  continuum~\cite{Lewandowski2019,Stauber2020,Novelli2020}.

Recently a new graphene/$\alpha-$RuCl$_3$ heterostructure has attracted significant attention~\cite{Zhou2019,Mashhadi2019,wang2020modulation,Basov2020}, since the Mott insulating $\alpha-$RuCl$_3$ layer is a promising candidate for realizing the seminal Kitaev quantum spin liquid (QSL)~\cite{Kitaev2006,Rau2016,Winter2017,Knolle2018,Takagi2019}. The quasi-2D material $\alpha-$RuCl$_3$ has long-ranged magnetism at low temperatures due to additional interactions beyond the  bond oriented Kitaev exchange but is believed to be in proximity to a QSL phase~\cite{banerjee2016proximate}. 
The lattice-mismatch between graphene and $\alpha-$RuCl$_3$ induces strain which has been shown to enhance the Kitaev spin exchange~\cite{Biswas2019,Gerber2020} bringing the system closer to the Kitaev QSL with its fractionalized Majorana fermion excitations. However, the graphene layer is also strongly affected because of a charge transfer from the itinerant to the insulating layer as observed experimentally~\cite{Mashhadi2019,wang2020modulation,Basov2020} and in accordance with ab-initio calculations~\cite{Biswas2019}. Graphene becomes hole-doped and $\alpha-$RuCl$_3$ electron-doped with the Fermi energy lying within the correlated narrow Ru-band which is almost flat in the Brillouin zone except for a small hybridization region~\cite{Biswas2019}.  
Recent experiments have observed plasmons in graphene/$\alpha-$RuCl$_3$ heterostructures with an excess damping mechanism attributed to the correlated insulating layer~\cite{Basov2020}. However, it is an outstanding question how the unusual excitations are linked to correlation effects of  $\alpha-$RuCl$_3$? More generally, it has remained unexplored whether plasmonic excitations can be used to probe correlation effects related to QSL fluctuations in correlated heterostructures?

\begin{figure}[tp]
	\includegraphics[width=\linewidth]{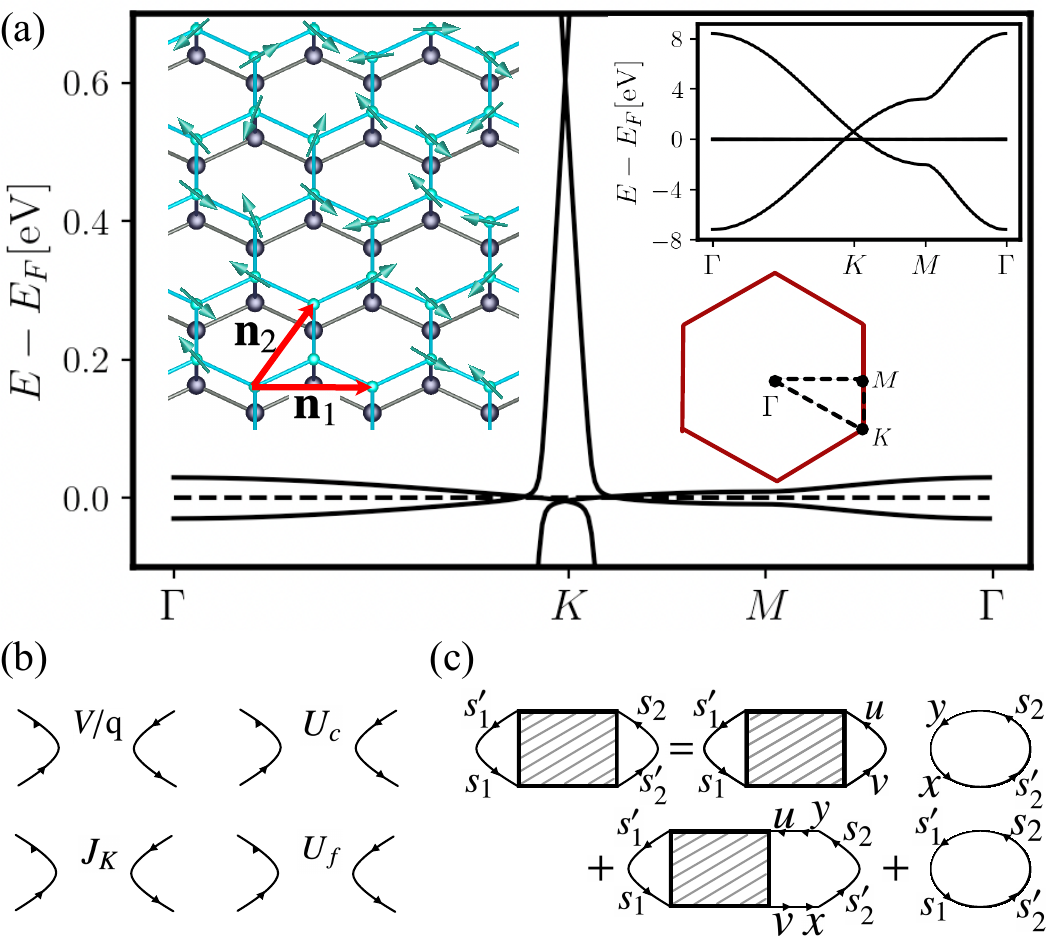}
	\caption{(a) The band structures of the effective model Eq.~\eqref{eq:4bandsh0} with $u=0.6$ eV, $t=2.6$ eV, $J=0.1$ eV, and $K=0.01$ eV~\cite{leeb2020}.  The dashed line is the Fermi energy $E_F=0$. \emph{The left inset plot} shows the schematic picture of the Kitaev-graphene lattice and corresponding lattice vectors $\bm{n}_{1}$ and $\bm{n}_{2}$. The upper layer with $S=1/2$ spins depicts for Kitaev Mott insulator and the other one the itinerant graphene layer. \emph{The right inset plots} shows the whole band structure and the Brillouin zone of honeycomb lattice. (b) Diagrammatic representations of the Coulomb interaction $V/{q}$, on-site Hubbard repulsion $U_c(f)$ and $U_f$, and Kondo couplings $J_K$. (c) The diagrammatic representation of the Dyson equation for the RPA charge susceptibility. }\label{fig:bands}
\end{figure}

In this work, we show that the interplay of correlated flat bands and strong local interactions can lead to novel plasmon excitations. We develop a microscopic theory of collective charge excitations in a minimal bilayer Kondo-Kitaev lattice model of $\alpha-$RuCl$_3$ on top of graphene with interlayer spin-only Kondo couplings. The Kondo-Kitaev model has a rich phase diagram displaying a fractionalized Fermi liquid, $p$-wave superconductivity, and heavy Fermi-liquid (hFL) phase as calculated within a self-consistent Abrikosov fermion mean-field theory~\cite{Kim2018,Vojta2018}. 
In the hFL phase, the fractionalized fermions of the Kitaev QSL acquire charge by hybridizing with the itinerant electrons from graphene which results in an almost flat Dirac band at the Fermi energy whose bandwidth is set by the Kitaev exchange and a hybridization by the local Kondo coupling, see Fig.\ref{fig:bands}~(a). Recently, it was shown that ab-initio calculations and experimental constraints can be used to determine the microscopic parameters of the effective low energy hFL band structure, which has been employed to explain the non-Lifshitz Kosevich temperature dependence of quantum oscillations measured in graphene/$\alpha$-RuCl$_3$ heterostructures~\cite{leeb2020}. 

Here, we calculate the dynamical charge susceptibility  for the hFL phase and study the plasmon dispersions over the Brillouin zone in both low- and high-energy scales taking into account the effect of different local interactions. As the main result, we find new plasmonic modes whose small momentum behavior is consistent with recent experiments on graphene/$\alpha$-RuCl$_3$ heterostructure~\cite{Basov2020}.

{\em Effective model of a Kitaev-Graphene system.---}\label{sec:model}
Our starting point is the Kondo-Kitaev lattice model with a ferromagnetic Kitaev layer in which the $S=1/2$ spins $\bm{S}_i$ are coupled to conduction electrons via the on-site antiferromagnetic Heisenberg Kondo coupling.
Within the framework of a parton theory, the $S=1/2$ spins can be  represented as bilinear forms of Abrikosov fermions, $\bm{S}_i=\frac{1}{2}f^\dagger_{i\sigma}\bm{\tau}_{\sigma\sigma'}f_{i,\sigma'}$, where $\bm{\tau}=(\tau^x,\tau^y,\tau^z)$ are three Pauli matrix and the summation over repeated spin indices $\sigma$'s is assumed. This representation enlarges the Hilbert space and a local constraint $\sum_{\sigma}f^\dagger_{i\sigma}f_{i\sigma}=1$ has to be imposed to restore the physical Hilbert space of spin-1/2's. Within a self-consistent parton mean-field solution, a hFL phase is realized in a large part of the phase diagram~\cite{Kim2018,Vojta2018} described by a quadratic Hamiltonian $H_0$, which has recently been shown to capture the essential aspects of the graphene/$\alpha-$RuCl$_3$ electronic structure~\cite{leeb2020}. 
In momentum space, it is expressed in terms of itinerant electrons $c_{s,\bm{k},\sigma}$ and Abrikosov fermions $f_{s,\bm{k},\sigma}$ ($s=1,2$) as~\cite{Vojta2018,leeb2020}
\begin{align}
H_0=\sum_{\sigma,\bm{k}}
\left(\begin{array}{l}c_{1,\bm{k},\sigma}\\c_{2,\bm{k},\sigma}\\f_{1,\bm{k},\sigma}\\f_{2,\bm{k},\sigma}\end{array}\right)^\dagger
\left(\begin{array}{cccc}
W & t\theta^*_{\bm{k}} &  J/2  & 0 \\
t\theta_{\bm{k}} & W & 0  & J/2 \\
J/2 & 0 & 0 &   K\theta^*_{\bm{k}} \\
0& J/2 &  K\theta_{\bm{k}}  &  0 \\
\end{array}\right)
\left(\begin{array}{l}c_{1,\bm{k},\sigma}\\c_{2,\bm{k},\sigma}\\f_{1,\bm{k},\sigma}\\f_{2,\bm{k},\sigma}\end{array}\right),\label{eq:4bandsh0}
\end{align}
where $\theta_{\bm{k}}=1+\exp(i\bm{k}\cdot\bm{n}_1)+\exp(i\bm{k}\cdot\bm{n}_2)$, $t$ ($K$) is the band parameter of $c$- ($f$-) fermions , $J$ is the hybridization strength, and $W$ is the energy shift between the graphene Dirac cone of $c$-fermions and the Kitaev Dirac cone of $f$-fermions. Note that the low energy scales $K, J$ are set by the Kitaev and Kondo exchange. Throughout this work, we consider the regime $t\gg{}K$ and $t>J$, where the Kitaev Dirac bands are almost flat. For convenience, we also adapt the notation of  $c_{3(4),\bm{k},\sigma}\equiv{}f_{1(2),\bm{k},\sigma}.$
The characteristic energy spectrum of $H_{0}$ is shown in Fig.~\ref{fig:bands}~(a). The hopping parameter is fixed as $t=2.6$ eV to adapt the slope of the graphene Dirac cone and the large energy shift $W\approx0.6$ eV is in accordance with the charge transfer from graphene to $\alpha$-RuCl$_3$~\cite{Biswas2019}. 
The Fermi energy ($E_F=0$) lies with the two flat Kitaev Dirac bands.

We are interested in the interaction induced fluctuations on top of this effective electronic structure and, therefore, concentrate on the following Hamiltonian
\begin{align}
	H_{\text{hetero}}=H_{0} + H^{(f)}_{\text{U}} + H^{(c)}_{\text{U}} + H_{\text{C}} + H_{\text{K}},  \label{eq:Hhetero}
\end{align}
 where the last four terms are quartic interaction terms.
As mentioned above, the Abrikosov fermions have a local  singly-occupancy constraint which can be effectively imposed by an on-site Hubbard term $H^{(f)}_{\text{U}}=U_f/2\sum_{s,\bm{r},\sigma}\left(f^{\dagger}_{s,\bm{r},\sigma}f_{s,\bm{r},{\sigma}}f^{\dagger}_{s,\bm{r},-\sigma}f_{s,\bm{r},-\sigma}\right)$, where the convention that $\sigma=1~(-1)$ for spin $\uparrow$ ($\downarrow$) has been adapted.
In practice, we will keep the on-site Hubbard $U_f$ very large but finite to enforce the constraint of $f-$fermions.
$H^{(c)}_{\text{U}}$ is an additional Hubbard term for $c$-fermions parameterized by repulsive strength $U_c$.
The fourth term, $H_{\text{C}}$, is the Coulomb interaction and reads
$H_{\text{C}}=\frac{1}{\mathcal{N}}\sum_{s,s'=1}^4\sum_{\bm{k,k',p},\sigma,\sigma'}\frac{V}{{q}}e^{-\bm{q}\cdot{}\bm{r}_{ss'}}\left(c^{\dagger}_{s,\bm{k+q},\sigma}c_{s,\bm{k},{\sigma}}c^{\dagger}_{s',\bm{k'-q},{\sigma}'}c_{s',\bm{k'},{\sigma}'}\right)$,
where $\mathcal{N}$ denotes the number of total sites, $q\equiv|\bm{q}|$, and $\bm{r}_{ss'}=0$ [$(\bm{n}_1+\bm{n}_2)/3$] if $s$ and $s'$ are on the same (different) sublattice(s).
Finally, $H_{\text{K}}$ is the Kondo coupling of strength $J_K$  between localized $c$- and $f$-fermions, which reads $H_{\text{K}}=J_K/4\sum_{s,\bm{r}}\left(c^\dagger_{s,\bm{r},\sigma_1}\bm{\tau}_{\sigma_1\sigma_1'}c_{s,\bm{r},\sigma_1'}\right)\cdot{}\left(f^\dagger_{s,\bm{r},\sigma_2}\bm{\tau}_{\sigma_2\sigma_2'}f_{s,\bm{r},\sigma_2'}\right)$. 
We note that in principle, the hybridization strength $J$ and the Kondo coupling $J_K$ only differ by a renormalized mean-field parameter~\cite{Kim2018,Vojta2018,leeb2020}, but here we treat $J_K$ as an independent parameter to investigate its qualitative effect on excitations. 

\begin{figure*}[htp]
	\includegraphics[width=\linewidth]{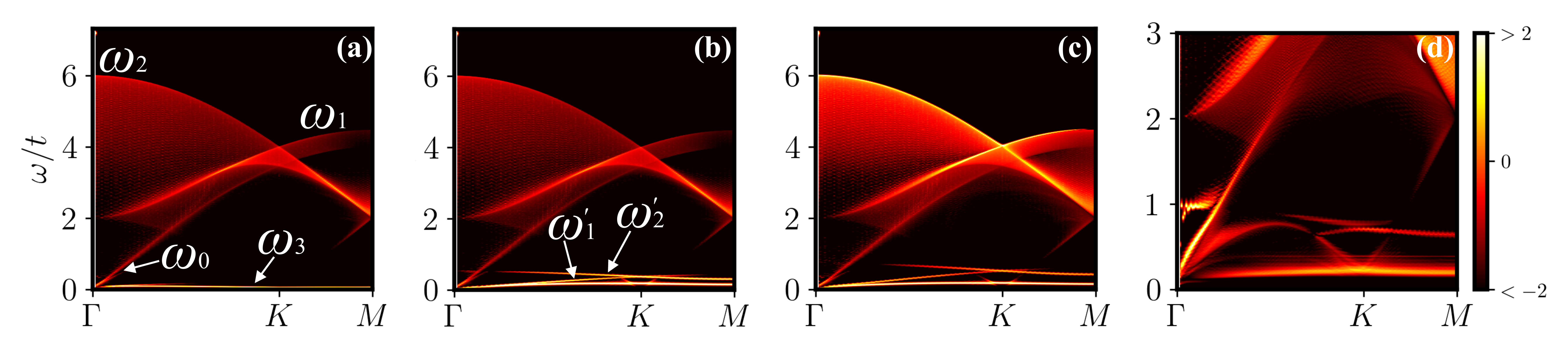}
	\caption{  The RPA energy loss function $1/{\rm Im}[\epsilon^{\text{RPA}}(\bm{q},\omega)]$ on a log scale along the $\bm{\Gamma-K-M}$ direction for parameters $J=0.1$ eV,  $K=0.01$ eV, $V=6$ eV, and (a) $U_f=U_c=J_K=0$ eV, (b) $U_f=120$ eV, $U_c=J_K=0$ eV, and (c) $U_f=240$ eV, $U_c=J_K=3$ eV.
	 (d) $1/{\rm Im}[\epsilon^{\text{RPA}}(\bm{q},\omega)]$ on a log scale with finite sublattice-symmetry-breaking term of  $\sum_{s}(-1)^sf^\dagger_{s,\bm{k},\sigma}f_{s,\bm{k},\sigma}$ for $J=0.1$ eV,  $K=0.01$ eV, $V=3$ eV, $U_f=120$ eV, and $U_c=J_K=3$ eV. 
     For numerical purposes we set the damping constant $\delta\approx{}0.01$ eV~\cite{appendix}.}\label{fig:rpaepsilon}
\end{figure*}

{\em Plasmons within a Random Phase Approximation.---}\label{sec:plasmon}
As a collective oscillating charge density mode, a plasmon is described by the total response of the systems to external potentials. 
Thus, it can be characterized by the total dynamical charge correlation function defined as 
\begin{align}
P(\bm{q},\omega)=\sum_{s,s'=1}^{4}\sum_{\sigma\sigma'}\int_{0}^{\beta}d\tau{}e^{i\omega\tau}\langle{}T_{\tau}\rho_{s\sigma}(\bm{q},\tau)\rho_{s'\sigma'}(-\bm{q},0)\rangle,\label{eq:ddmat}
\end{align}
where $\rho_{s\sigma}(\bm{q})=\frac{1}{\mathcal{N}}\sum_{\bm{k}}c^\dagger_{s,\bm{k},\sigma}c_{s,\bm{k+q},\sigma}$ is the total density operator ($s=1,...,4$).
In order to account for all kinds of interactions in Eq.~\eqref{eq:Hhetero}, we define the bare dynamical charge susceptibility tensor as  
\begin{align}
\begin{split}
&[\chi^0(\bm{q},\omega)]^{s_1s_1'}_{s_2s_2'}=\frac{1}{\mathcal{N}^2}\sum_{\bm{k}\bm{k}'\sigma\sigma'}\int_{0}^{\beta}d\tau{}e^{i\omega\tau}\\&\langle{}T_{\tau}c^\dagger_{s_1,\bm{k},\sigma}(\tau)c_{s_1',\bm{k+q},\sigma}(\tau)c^\dagger_{s_2,\bm{k}',\sigma'}(0)c_{s_2',\bm{k'-q},\sigma'}(0)\rangle_0,\\
\end{split}
\end{align}
where $\langle...\rangle_0$ is a canonical ensemble average with respect to the bare Hamiltonian Eq.~\eqref{eq:4bandsh0}. 
Up to the zeroth order in interactions, the bare charge correlation function reads 
$P^0(\bm{q},\omega)=\sum_{s,s'=1}^{4}[\chi^0(\bm{q},\omega)]^{ss}_{s's'}$.
The interactions then need to be treated self-consistently, leading to the following total dynamical dielectric function
$\epsilon(\bm{q},\omega) = 	P(\bm{q},\omega)/P^{0}(\bm{q},\omega)$ with plasmon excitations given by the zeros of the energy loss function, {\em e.g.}, the inverse of the imaginary part of $\epsilon(\bm{q},\omega)$.

For analyzing the effect of the different interaction channels we treat the dynamical correlation functions within the random phase approximation (RPA). 
For graphene, this is well justified for $q\ll{}k_F$ with $k_F$ the Fermi wavenumber.
Plasmon dispersions are determined by the zeros of  imaginary part of the RPA dielectric function as
$\epsilon^{\text{RPA}}(\bm{q},\omega)=P^{\text{RPA}}(\bm{q},\omega)/P^{\text{0}}(\bm{q},\omega)$
where $P^{\text{RPA}}(\bm{q},\omega)\equiv{}\sum_{ss'}[\chi^{\text{RPA}}(\bm{q},\omega)]^{ss}_{s's'}$.
The RPA charge susceptibility tensor $\chi^{\text{RPA}}(\bm{q},\omega)]^{s_1s_2}_{s_1's_2'}$ is obtained via a generalized Dyson equation~\cite{schrieffer1989,Chubukov1992,Knolle2010RPA} given by
\begin{align}
\begin{split}
&[\chi^{\text{RPA}}]^{s_1s_1'}_{s_2s_2'} = [\chi^{0}]^{s_1s_1'}_{s_2s_2'} + [\chi^{\text{RPA}}]^{s_1s_1'}_{uv}[V^{{l}}-V^{{b}}]^{uv}_{xy}[\chi^{0}]^{xy}_{s_2s_2'},\label{eq:RPA}	\end{split}
\end{align}
where $V^{b}$ ($V^{l}$) is a vertex for bubble (ladder) diagrams and repeated indices $u,v,x,y$ are summed over.
The Dyson equation is schematically shown in Fig.~\ref{fig:bands} (c).
All the tensors in Eq.~\eqref{eq:RPA} can be treated as matrices with row index $s_1s_1'$ and  column index $s_2s_2'$ and then the solution of Eq.~\eqref{eq:RPA}  in matrix form is $\chi^{\text{RPA}} = \chi^{0}[1- (V^{{l}}-V^{{b}})\chi^{0}]^{-1}$.
The spin indices of the vertices $V^{{b(l)}}$ vanish in Eq.\eqref{eq:RPA} because we have summed over all spin degrees of freedom.
The nonzero elements of the bubble contribution are
$[V^{{b}}]^{11(33)}_{11(33)}=[V^{{b}}]^{22(44)}_{22(44)}=\frac{4V}{{q}}+U_c(U_f)$,  $[V^{{b}}]^{11}_{33}=[V^{{b}}]^{22}_{44}=\frac{4V}{{q}}$,
$[V^{{b}}]^{11}_{22(44)}=[V^{{b}}]^{33}_{22(44)}=\frac{4V}{{q}}e^{-\bm{q}\cdot(\bm{n}_1+\bm{n}_2)/3}$ and we have $[V^{{b}}]^{aa}_{cc}=[V^{{b}}]^{cc}_{aa}$.
When calculating the ladder diagrams, we ignore the contributions from $\bm{q}$-dependent Coulomb interactions which carry momentum-transfer processes. 
Consequently, the remaining nonzero elements of the ladder vertex read 
$[V^{{l}}]^{13}_{31}=[V^{{l}}]^{31}_{13}=[V^{{l}}]^{24}_{42}=[V^{{l}}]^{42}_{24}=J_K$.
Notice that the contributions from on-site Hubbard repulsion $U_c~(U_f)$ for ladder diagrams and from Kondo coupling $J_K$ for bubble diagrams are zero after summing over all spin indices.

{\em Numerical results.---}\label{sec:numresluts}
For a realistic graphene/$\alpha$-RuCl$_3$ system, we follow earlier work~\cite{leeb2020} and fix $W=0.6$ eV, $t=2.6$ eV, and $E_F=0$ eV to numerically compute $\chi^{(0)}$ and $\chi^{\text{RPA}}$. In general, we are interested in collective excitations over the whole Brillouin zone of the honeycomb model. 
The results of our calculations are presented in Fig.~\ref{fig:rpaepsilon} at zero temperature and for several values of the different interaction parameters.

The graphene subsystem hosts three plasmon bands induced by Coulomb interactions, \emph{e.g.}, one acoustic-like band $\omega_{0}$ as well as two optical bands $\omega_{1}$ and $\omega_{2}$, see Fig.~\ref{fig:rpaepsilon}~(a). 
Similar plasmon dispersions in pure  graphene systems have been studied beyond the  $\bm{k}\cdot\bm{p}$ approximation~\cite{Stauber2010_1,Stauber2010_2,hill2009}.
As the graphene Dirac cone at $\mu=0.6$ eV is far away from the Fermi energy,  $\omega_{0}$ is dispersing as $\propto{}{q}$ rather than $\propto\sqrt{{q}}$ at lower frequencies and plunges into a p-e continuum at higher frequencies.
The two optical bands $\omega_{1}$ and $\omega_{2}$ are degenerate at the $\bm{K}$ point and form a big crossing in the high-symmetry direction $\bm{\Gamma-K-M}$.  
$\omega_{2}$ hosts co-called $\pi$ plasmons associated with the Van Hove singularity at the $\bm{M}$ point~\cite{Eberlein2008,Kinyanjui2012,Stauber2010_1}.

In addition to the three graphene plasmon bands, we find a novel low energy plasmon, $\omega_{3}\sim{}J$, see Fig.~\ref{fig:rpaepsilon} (a). Note that for illustrative purposes, we have chosen a much larger value of $J\sim{}0.1$ eV than determined in Ref.~\cite{leeb2020}.  The flat plasmon band originates from the flat electronic bands of the Kitaev layer and barely changes as the interaction couplings vary.
However, a finite on-site repulsion $U_f$ of the formerly Mott insulating layer leads to two extra plasmon branches above $\omega_{3}$, namely $\omega_{1}'$ and $\omega_{2}'$, forming a smaller and flatter crossing in the $\bm{\Gamma-K-M}$ direction with an intersection at $\bm{K}$ point, as shown in Fig.~\ref{fig:rpaepsilon} (b) and (c). 
This flatter crossing at low energies only depends on $U_f$. It is enlarged and pushed to higher energy region as $U_f$ increases, see Fig.~\ref{fig:rpaepsilon}(c)~\footnote{ For illustration purposes and imposing the local constraint of Abrikosov fermions, we have used very large Hubbard repulsion to demonstrate $\omega'_1$ and $\omega'_2$. These two bands do not qualitatively change as $U_f$ increases, and the higher order corrections~\cite{Sodemann2012} must be considered to quantitatively determine  $\omega'_1$ and $\omega'_2$. }.
Finally, we find that the main effect of moderate Kondo and Hubbard interactions, $J_K$ and $U_c$, is to enhance the signals of $\omega_{1}$ and $\omega_{3}$ [see Fig.~\ref{fig:rpaepsilon} (c)].

We note that there exist a temperature scale above which a crossover from the hFL phase to the decoupled phase occurs~\cite{Vojta2018}. In the decoupled phase, the hybridization strength renormalizes to zero and consequently the flat plasmon band $\omega_3$ and the two emergent bands $\omega'_1$ and $\omega'_2$ all disappear. More details about finite temperature effects and disorder broadening can be found in the Supplemental Material~\cite{appendix}.

An experimentally important aspect so far not accounted for in our minimal model is the lattice mismatch between the two layers, which is expected to gap the flat Dirac cone of the Kitaev layer around $E_F$. 
This can be effectively mimicked by introducing a ``$\tau_z$'' term of $\sum_{s=1,2}(-1)^sf^\dagger_{s,\bm{k},\sigma}f_{s,\bm{k},\sigma}$ which breaks the sublattice symmetry for the Kitaev layer. We find that a finite and small ``$\tau_z$'' term does not change our main results. However, it pushes the flat plasmon band $\omega_{3}$ to higher energy and opens a small gap at the flat crossing around the $\bm{K}$ point, see Fig.~\ref{fig:rpaepsilon} (d). A similar effect on plasmons due to a correlation-driven sublattice asymmetry has been discussed recently in connection with TBG~\cite{Fahimniya2020}. The sublattice asymmetry endows the plasmons with a dipole moment which is expected to lead to a stronger experimental response.  

\begin{figure}[]
	\includegraphics[width=\linewidth]{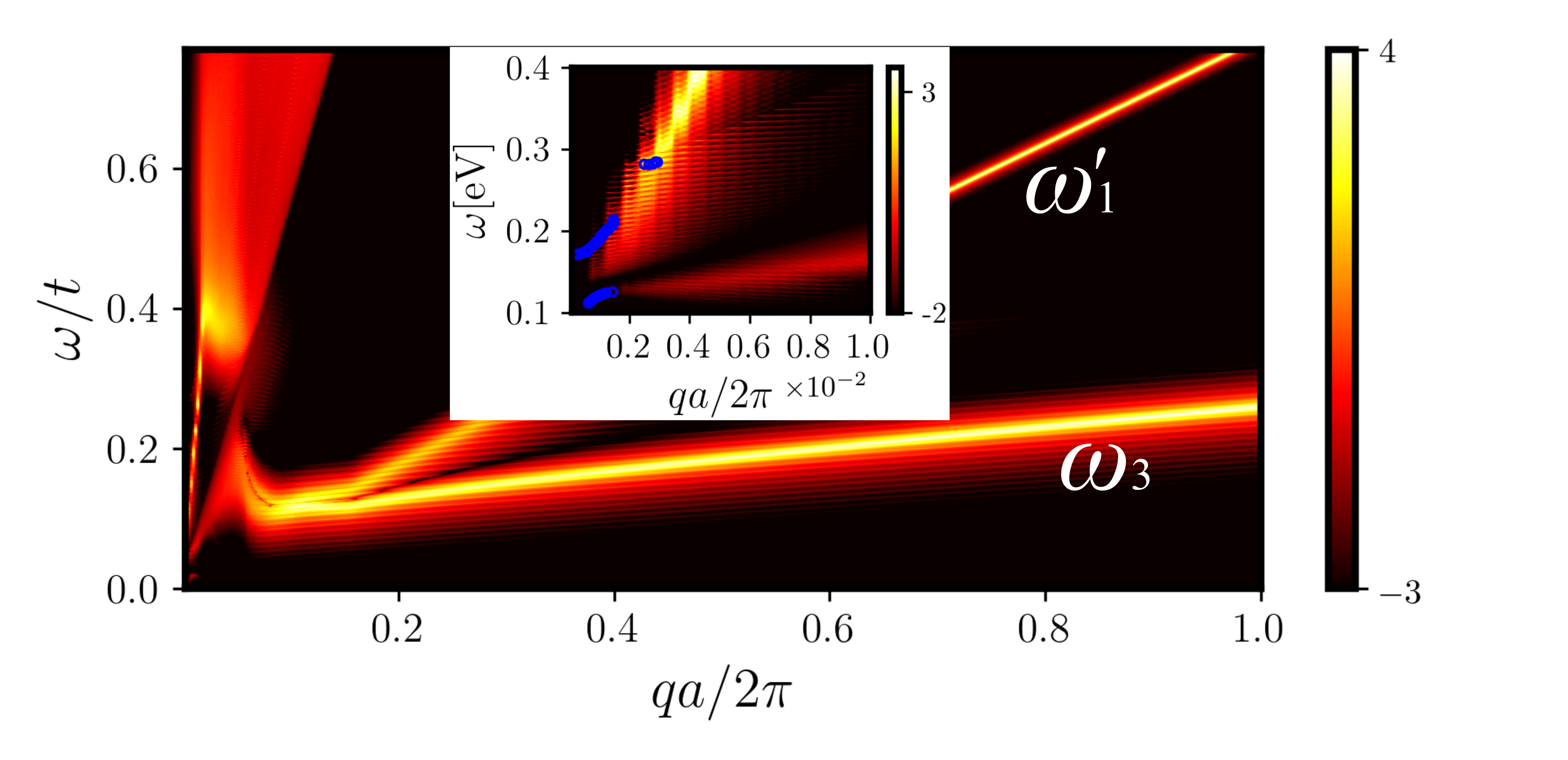}
	\caption{ The imaginary part of the inverse RPA dielectric functions $1/{\rm Im}[\epsilon^{\text{RPA}}(\bm{q},\omega)]$ of the linearized Hamiltonian in a log scale with $K=0.05$ eV, $J=0.12$ eV, $V=3$ eV, $U_c=J_K=1$ eV, and $U_f=20$ eV. Inset: Comparison of the plasmon dispersions from experimental data (blue circles) and the linearized low energy model Eq.~\eqref{eq:Hhetero}. The model parameters are the same as those in the main plot. The experimental data are collected from Fig. 2 in Ref.~\cite{Basov2020}. }\label{fig:expdata}
\end{figure}

{\em Comparison to experiment.---}
In experiments, the scattering-scanning near-field optical microscopy (s-SNOM) method~\cite{Atkin2012,Basov2016,Low2017,Sunku2018,Ni2015,Hu2017} can be used to measure the dispersions of collective charge modes, for example, plasmons in (twisted bilayer) graphene~\cite{Basov2016,Low2017,Woessner2015,Hesp2019}.
In Ref.~\cite{Basov2020}, the authors recently performed s-SNOM experiments on the new graphene/$\alpha$-RuCl$_3$ heterostructures on a SiO$_2$/Si substrate encapsulated with hexagonal boron nitride (hBN) and extract the dispersions for plasmons. 
The experimental data resolves two long-wavelength plasmon dispersions shown as blue circles in Fig.~\ref{fig:expdata}: a lower branch in the region of $\omega=0.111\sim0.136$ eV and $qa=0.004\sim{}0.0094$ and an upper branch spanning the region of $\omega=0.171\sim0.285$ eV and $qa=0.0018\sim{}0.025$. 
Here, $a\approx0.25$ nm is the lattice constant of graphene. 
These two plasmon bands are separated by a region of SiO$_2$ and hBN phonons~\cite{Dai2015,Basov2020}. Therefore, it was argued that the experimental response can be well explained by the interplay of surface plasmon polaritons of doped graphene and the hyperbolic phonon polaritons in hBN~\cite{wu2015,Dai2015,Hwang2010}. However, the unusually large damping measured for these modes was an indication of potential  correlation effects from the $\alpha$-RuCl$_3$ layer~\cite{Basov2020}. 

Alternatively, we can compare our flat and correlated plasmon bands of the correlation driven hFL phase with the experimental data on the graphene/$\alpha$-RuCl$_3$ interface~\cite{Basov2020}. 
Since we are now interested in the long-wavelength limit with ${q}\ll{}k_F$, we expand the terms of $\theta_{\bm{K}+\bm{k}}$ in Eq.~\eqref{eq:4bandsh0} around momentum $\bm{K}$ to obtain a $\bm{k}\cdot\bm{p}$ Hamiltonian $\tilde{H}_{0}$. 
The plasmon bands $\omega_{0}$, $\omega_{3}$, and $\omega'_{1}$ from $\tilde{H}_{0}$ are shown in Fig.~\ref{fig:expdata}.  
It turns out that 
$\omega_{0}$ is gapped at $q=0$ which differs from the usual Dirac plasmons and
$\omega_{3}$ merges with  $\omega'_{1}$ around the momentum point where $\omega_{0}$ becomes damped.
Notice that $\omega_{3}$ is not an exactly flat band at large $q$ anymore but has a finite slope due to the $\bm{k}\cdot\bm{p}$ approximation. 
In the inset panel of Fig.~\ref{fig:expdata}, we compare our dispersion with the experimental measurements. 
Intriguingly, it reproduces the available data for the upper plasmon band with $\omega_{0}$, and the other lower band roughly matches the tail of the $\omega_{3}$ and/or $\omega'_{1}$ branch originating from the correlated Kitaev layer. 

{\em Discussion and summary.---} 
We have analyzed the charge response of graphene/$\alpha$-RuCl$_3$ heterostructures within a minimal model. At low temperatures, the Kondo-Kitaev lattice leads to a peculiar electronic structure of a hFL with the formerly fractionalized excitations of the correlated Kitaev layer hybridized with the Dirac electrons. Within an RPA treatment, we investigated the effect of various interactions on the dynamical charge susceptibility, \emph{e.g}, Coulomb interaction, on-site Hubbard repulsion for both layers, and interlayer Kondo coupling.
We found a novel low energy branch of flat plasmon bands over the entire Brillouin zone which originates from the Kitaev layer. A large Hubbard repulsion which is generically expected because of the Mott insulating nature of the $\alpha$-RuCl$_3$ film leads  to two correlated optical plasmon branches above the flat plasmon band which look like a zoomed-out version of two optical plasmon bands of the doped graphene layer at much higher energy.

From a linearized Hamiltonian, we examined the plasmons in the low energy limit with ${q}\ll{}k_F$ and argue that our theory is consistent with the recent experimental data on graphene/$\alpha$-RuCl$_3$ heterostructures~\cite{Basov2020}. It would be desirable to extend the experimental measurements to larger momenta which could directly verify our predictions of flat and correlated plasmon bands with the potential to  shed new light on the proximate QSL of $\alpha$-RuCl$_3$. Similarly, we expect that signatures of the hFL will be visible in scanning tunneling microscopy and photo-emission spectroscopy. 

In general, we showed that the collective charge response provides a direct probe of correlation effects in heterostructures, {\it e.g.}, for understanding the interplay of fractionalized excitations and itinerant electrons. We expect our theory to be applicable in other systems like Bi$_2$Se$_3$ grown on $\alpha$-RuCl$_3$~\cite{Park2020}, the Dirac hFL of graphene intercalated with Cerium~\cite{hwang2018emergence} or in quantum many-body phases of TBG. 

{\em Acknowledgments.---}
We acknowledge helpful discussion and collaboration on related work with R. Valenti, M. Burghard, K. Burch and V. Leeb. 
H.-K. Jin is funded by the European Research Council (ERC) under the European Unions Horizon 2020 research and innovation program (grant agreement No. 771537).

\bibliography{PlasmonKitaevGraphene.bib}

\begin{widetext}
\begin{center}
{\bf Supplemental material for ``Flat and Correlated Plasmon Bands in Graphene/$\alpha$-RuCl$_3$ Heterostructures''}
\end{center}

	\setcounter{equation}{0}
	\setcounter{figure}{0}
	\setcounter{table}{0}
	
	\renewcommand{\theequation}{S\arabic{equation}}
	\renewcommand{\thefigure}{S\arabic{figure}}
	\renewcommand{\thetable}{S\Roman{table}}
	
	In this Supplemental Material, we discuss disorder and finite temperature effects on the plasmon bands in the graphene/$\alpha$-RuCl$_3$ heterostructures.
	
	\begin{figure}[ht]
		\includegraphics[width=\linewidth]{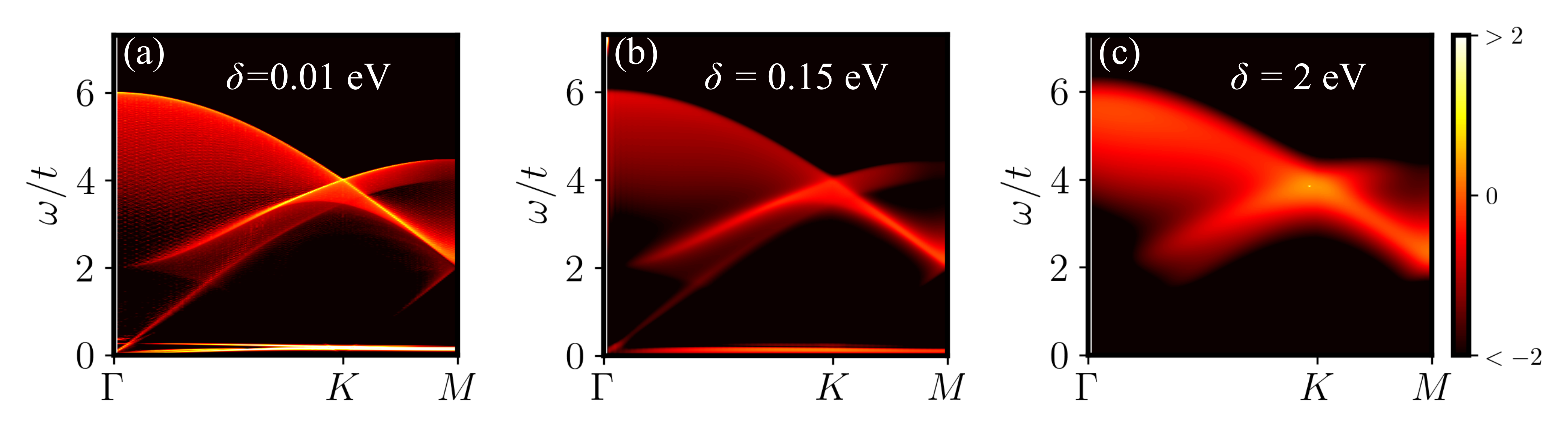}
		\caption{  
			The energy loss function (the imaginary part of the inverse  dielectric function ) $1/{\rm Im}[\epsilon^{\text{RPA}}(\bm{q},\omega)]$ on a log scale along the $\Gamma-K-M$ direction for parameters of $J=0.1$ eV, $K=0.1$ eV, $V=6$ eV, $U_c=J_K=0$ eV, and $U_f=20$ eV with variant damping constants $\delta$. }\label{fig:disorder}
	\end{figure}
	\section{Disorder effects}
	In graphene/$\alpha$-RuCl3 heterostructures, the disorder induced by, for instance, impurities and crystal defects are inevitable. This disorder effect usually will impact the lifetime of plasmon excitations.
	Here, we discuss the qualitative impact of disorder which will effectively result in a very large damping constant $\delta$ when calculating the bare charge susceptibility tensor $\chi^0(\bm{q}, \omega)$ which explicitly reads
	\begin{equation}
		\begin{split} 
			&[\chi^{(0)}(\bm{q},\omega)]^{s_1s_1'}_{s_2s_2'}=-\frac{2}{\mathcal{N}}\sum_{\bm{k}\mu\nu}\frac{n_f(\omega_{\mu,\bm{k}})-n_f(\omega_{\nu,\bm{k+q}})}{\omega_{\mu,\bm{k}}-\omega_{\nu,\bm{k+q}}+\omega+i\delta}[U_{\mu,\bm{k}}^{*}]_{s_1}[U_{\nu,\bm{k+q}}]_{s'_1}[U_{\nu,\bm{k+q}}^{*}]_{s_2}[U_{\mu,\bm{k}}]_{s'_2}.
		\end{split}
	\end{equation}
	Here $n_f$ is the Fermi function, $\omega_{\mu,\bm{k}}$ is the energy of the $\mu$-th band of quadratic Hamiltonian $H_0$ defined in the main text, and $U_{\mu,\bm{k}}$ is the four-component eigenvector associated with energy  $\omega_{\mu,\bm{k}}$.
	In Fig.~\ref{fig:disorder}, we plot the energy loss functions for different damping constants $\delta$.  
	In the main text, we usually set the damping constant $\delta\approx{}0.01$ eV.
	As we can see, a moderately large damping constant of $\delta=0.15$ eV broadens the peak of the energy loss function $1/{\rm Im}[\epsilon^{\text{RPA}}(\bm{q},\omega)]$, {\em i.e.}, weakening the signals of plasmon excitations.
	Crucially, the low-energy plasmons will be totally eliminated for strong disorder as shown via a large damping constant of $\delta\approx{}2$ eV, {\em e.g.}.

	\begin{figure}[t]
		\includegraphics[width=0.9\linewidth]{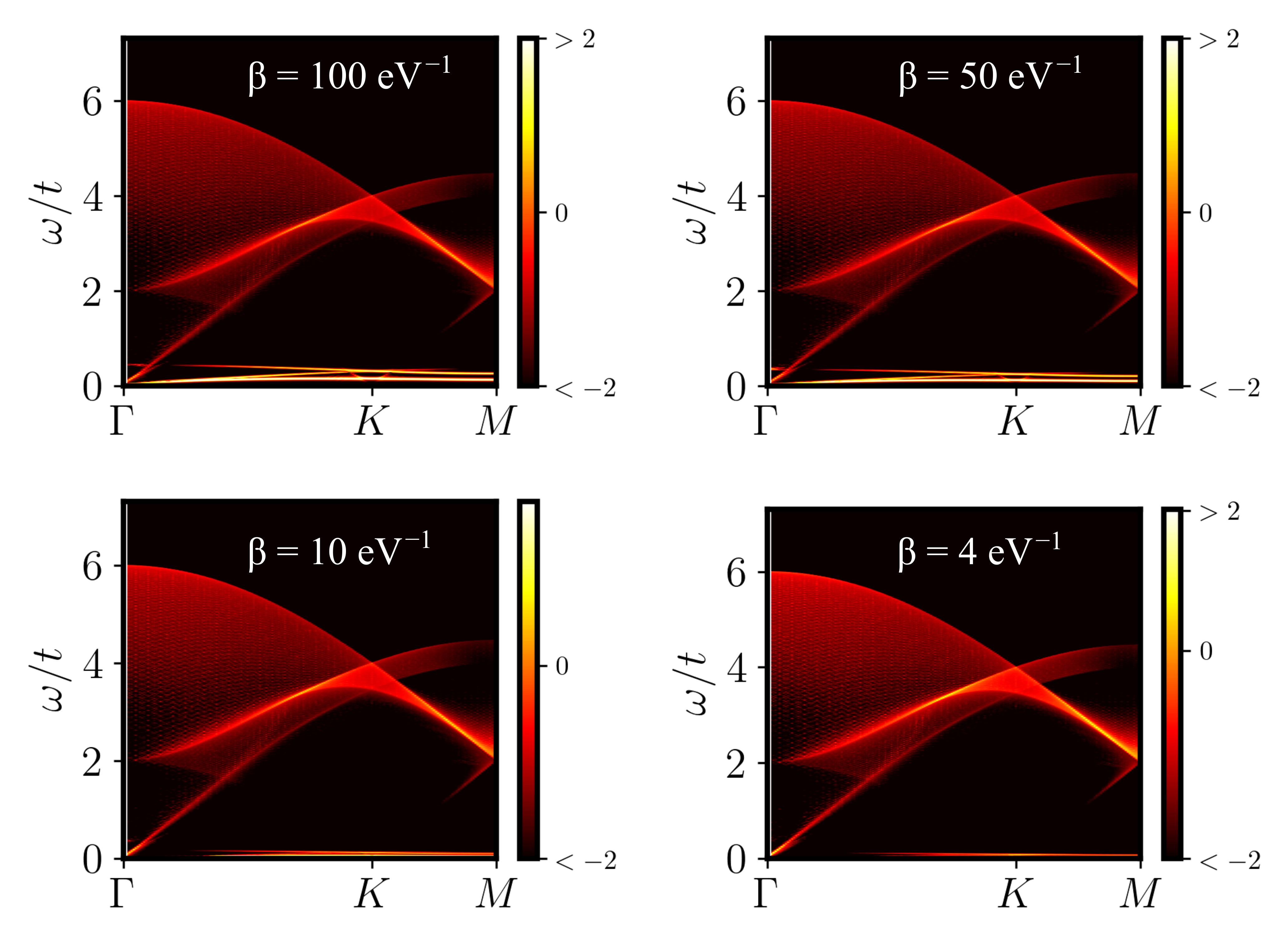}
		\caption{  
			The energy loss function (the imaginary part of the inverse  dielectric function ) $1/{\rm Im}[\epsilon^{\text{RPA}}(\bm{q},\omega)]$ on a log scale along the $\Gamma-K-M$ direction for parameters of $J=0.1$ eV, $K=0.1$ eV, $V=6$ eV, $U_c=J_K=0$ eV, and $U_f=20$ eV with different temperatures $\beta$. }\label{fig:finiteT}
	\end{figure}
	\section{Finite temperature effects}
	There are two qualitatively different effects of increasing temperature.
	First and most importantly, the finite temperature effect may destroy the heavy Fermi liquid phase, {\em i.e.}, the quadratic Hamiltonian $H_0$ in the main text with its effective hybridization which allows coupling to the insulating layer.
	There exists a temperature scale $\beta^\ast\approx4\ {\rm eV}^{-1}$ above which a crossover (manifest as a phase transition in the mean field treatment) from the heavy Fermi liquid phase to the decoupled phase occurs. 
	Notice that the transition temperature $\beta^*$ is just a rough estimate and more details can be found in Ref.~\cite{Vojta2018}.
	In the decoupled phase, the interlayer hybridization strength is effectively $J=0$ and consequently, the flat plasmon band $\omega_3$ and other two emergent bands $\omega'_1$ and $\omega'_2$ disappear.
	Second, within the heavy Fermi liquid phase, increasing temperature will lead to the standard broadening effects from the smearing of the Fermi function. 
	The plasmon bands for different temperatures are shown in Fig.~\ref{fig:finiteT}, where we have assumed that the interlayer hybridization strength $J$ does not change a lot as a function of temperature within the heavy Fermi liquid phase. 
	We find that finite temperature does not have a significant effect on the plasmon excitations. The flat plasmon band $\omega_3$ is still observable for $\beta>4$ eV$^{-1}$, whereas the emergent two plasmon bands $\omega'_1$ and $\omega'_2$ will gradually disappear around the critical temperature  $\beta^*\approx4\ {\rm eV}^{-1}$.

\end{widetext}

\end{document}